\documentclass[twocolumn,aps,jcp,amsmath,showpacs]{revtex4}
\usepackage{graphicx,dcolumn,bm,amssymb,amsmath}
\usepackage{subfigure}
\usepackage{color}
\usepackage[colorinlistoftodos]{todonotes}

\def\be{\begin{equation}}
\def\ee{\end{equation}}

\begin{document}

\title{Fragmentation and depolymerization of non-covalently bonded filaments}
\author{A. Zaccone$^{1}$, I. Terentjev$^{2}$, L. Di Michele$^{3}$ and E. M. Terentjev$^{3}$}
\affiliation{${}^1$Physics Department \& Institute for Advanced Study, Technische Universit\"{a}t M\"{u}nchen,
85748 Garching, Germany}
\affiliation{${}^2$Granta Design, 62 Clifton Rd, Cambridge CB1 7EG, U.K.}
\affiliation{${}^3$Cavendish Laboratory, University of Cambridge, JJ Thomson Avenue,
Cambridge CB3 0HE, U.K.}
\date{\today}
\begin{abstract}
\noindent  Protein molecules often self-assemble by means of non-covalent physical bonds to form extended filaments, such as amyloids, F-actin, intermediate filaments and many others. The kinetics of filament growth is limited by the disassembly rate, at which inter-protein bonds break due to the thermal motion. Existing models often assume that the thermal dissociation of subunits occurs uniformly along the filament, or even preferentially in the middle, while the well-known propensity of F-actin to depolymerize from one end is mediated by chemical factors (ADP complexation).  Here we show for a very general (and generic) model, using Brownian dynamics simulations and theory, that the breakup location along the filament is strongly controlled by the asymmetry of the binding force about the minimum, as well as by the bending stiffness of the filament. We provide the basic connection between the features of the interaction potential between subunits and the breakup topology. With central-force (that is, fully flexible) bonds the breakup rate is always maximum in the middle of the chain, whereas for semiflexible or stiff filaments this rate is either a minimum in the middle or flat. The emerging framework provides a unifying understanding of biopolymer fragmentation and depolymerization, and recovers earlier results in its different limits.
\end{abstract}
\maketitle

\section{INTRODUCTION}
The configuration-mediated directionality of non-covalent bonds between proteins explains their propensity to self-assemble into fibrils and filaments \cite{oosawa75,pollard86,dobson06,chandler05,irback03}. Protein filaments are ubiquitous in biology, forming inside the cells or in the extra-cellular matrix -- individually, in bundles, or in randomly crosslinked networks. They facilitate the propulsion in bacteria, they control the mechanical strength in cytoskeleton and the bending stiffness in axons, they allow positional control of organelles and provide the transport routes all around the cell \cite{oosawa75,dobson06,chandler05,irback03,niranjan03,adamcik10,tanaka06}. In a different situation, the self-assembly of proteins into amyloid fibrils impairs physiological activity and is the root cause of a number of organic dysfunctions \cite{dobson06,knowles09,fodera1,fodera2}. In yet another context, filaments are artificially or spontaneously assembled to achieve a specific function in the material, such as directed conductivity, plasmonic resonances, or just the mechanical strength in a fiber composite, with important technological applications \cite{colloid09,bonn98}. Finally, a conceptually related issue emerges in the denaturation of DNA~\cite{DiMicheleJACS}, for which the available theoretical framework \cite{peyrard89,mast13} cannot provide predictions about the topology of the disassembly process. The typical size of all these aggregates, and its time-evolution, are a non-trivial function of the rate at which bonds along the filament spontaneously dissociate due to the thermal motion of the assembled molecules. The dissociation rate and the distribution of fragments are important parameters which enter the master kinetic equation description of self-assembling filament size and populations.

A filament growth can be summarized by the reversible reaction: $A_n + A_1\rightleftharpoons A_{n+1}$, where the monomer subunit $A_1$ is added to an existing filament of $n$-units long. For the forward reaction, it is commonly accepted that association proceeds by the addition of a single subunit -- as opposed to the joining of larger segments -- because of the greater abundance of monomers with respect to active fragments. In contrast, despite the importance of thermal breakup in many fields of colloid science and technology~\cite{wu1,wu2}, its basic understanding is far from satisfactory. Several studies aimed to explain thermally-activated filament breakup in physical terms, came to the conclusion that fibrils of any respective size can aggregate, while the filament breakup can occur with equal probability anywhere along its length. In particular, Lee~\cite{lee09} has demonstrated that the thermal breakup occurs randomly along the chain, leading to daughter fragments of any size.  In yet another classical model based on equilibrium detailed-balance between the various aggregation and breakup events, by Hill~\cite{hill}, the highest breakup probability is for two fragments of equal size, i.e. the breakup rate is maximum in the middle.

Theoretical models in the past have focused on the simplified case of chains of harmonically bonded particles (subunits), so that the binding force is linear in the inter-protein displacement~\cite{hill,lee09}. In this approximation the normal modes of vibration of the chain are de-coupled, which makes the problem amenable to simpler analysis. Even in this case, previous theoretical models reached contradictory conclusions, with either flat breakup distribution or a pronounced maximum in the middle. However, the physical bonds linking protein filament subunits (such as hydrogen bonds and hydrophobic attraction) are strongly anharmonic. Then the problem becomes one of coupled nonlinear oscillators as in the famous Fermi-Pasta-Ulam problem~\cite{fermi55}, for which the typical vibration modes are no longer delocalized periodic waves but solitons~\cite{kruskal65}. This is also consistent with the finding~\cite{oliveira,vilgis} that in a strained Lennard-Jones chain, the strain is not uniformly distributed, but localized around the bond which is going to break first.
The standard tools of chemical dynamics and stochastic rate theory~\cite{zaccone12,haenggi90}, all based on the harmonic approximation and on normal modes, are therefore inapplicable~\cite{wiggins13,vilgis2}.

Here we develop a systematic microscopic understanding of this process based on Brownian dynamics simulation and theoretical arguments, {focusing on the {nonequilibrium} breakup phenomena. Hence we study the intrinsic breakup rates independent of any recombination phenomena which may occur at later stages leading eventually to an equilibrium size.} First of all, we discover that the topology of filament breakup critically depends on the bending stiffness of the chain.  Secondly, a clear connection is found between the anharmonicity of subunit interaction and the fragment distribution resulting from thermal breakup. The anharmonic Lennard-Jones or Morse-like binding potential in stiff or semiflexible filaments inevitably leads to a very strong preference for the breakup to occur at chain ends, but recover the uniform, flat fragment distribution in the limit of harmonic (or any other symmetric) potential. Importantly, it is not the bare anharmonicity which controls this effect, but, more precisely, the \textit{asymmetry} of the bonding potential about the minimum (larger force for bond compression than for extension), which is inherent to the most common anharmonic potentials. As we will show below, it is precisely the asymmetry in the potential which "breaks the symmetry" between dissociation rates at the middle of the filament and at the ends. Those rates are equal only for symmetric potentials like harmonic, and they always differ for asymmetric potentials.

In contrast, when the intermolecular interaction is purely of the central-force type, i.e. a fully flexible chain with no bending resistance, a bell-like distribution peaked in the middle is obtained in accord with the prediction of the Hill model. These findings can be understood with an argument based on counting the degrees of freedom per particle for the different potentials.
These results provide a fundamental link between the features of intermolecular interaction and the filament breakup rate and topology, and can be used in the future to predict, control and manipulate the filament length distribution in a variety of self-assembly processes in biological and nanomaterials applications.

\section{SIMULATIONS}
\begin{figure} 
\includegraphics[width=.42\textwidth]{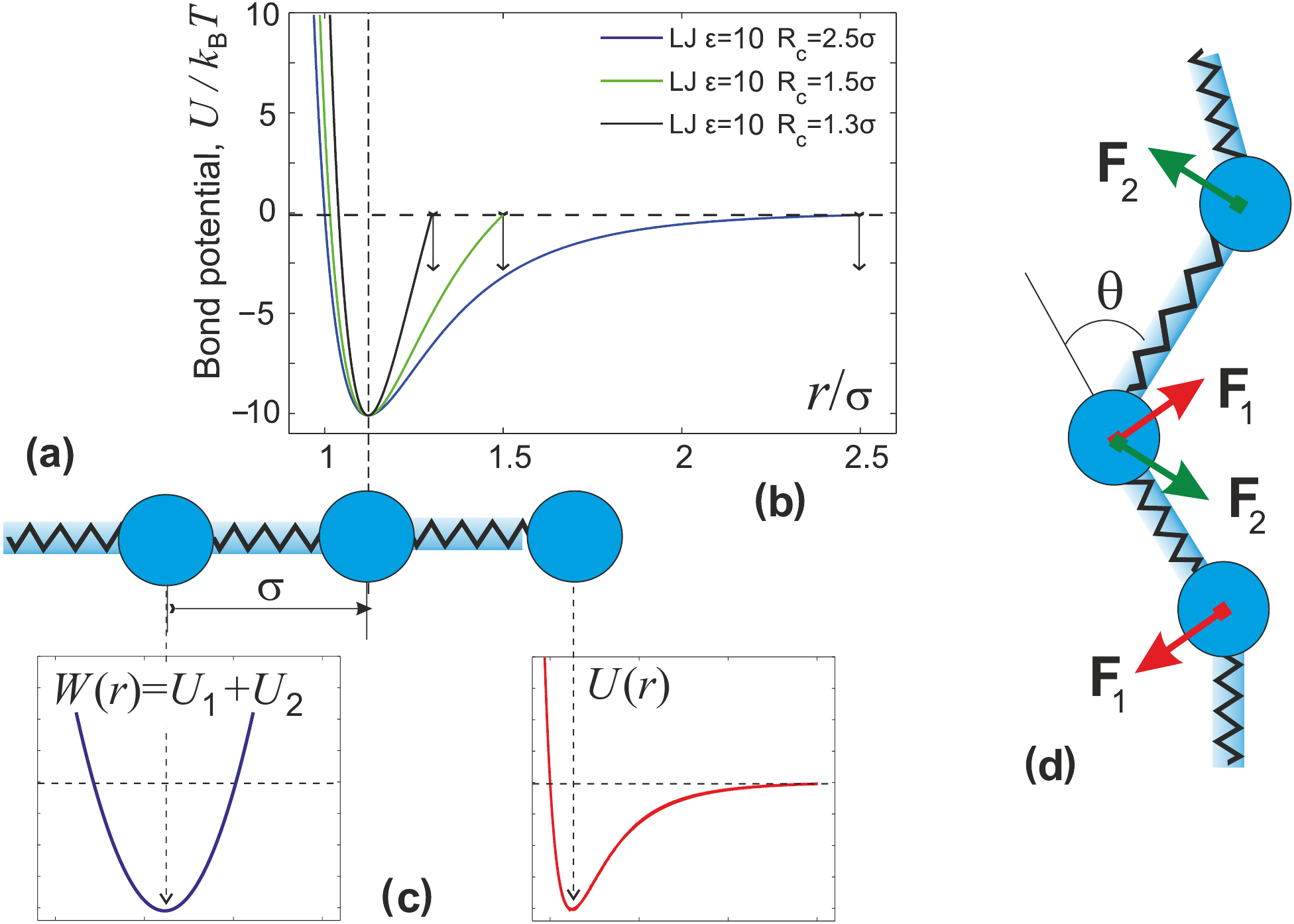}
\caption{(a) Scheme of the coarse-grained nanofilament as a sequence of subunits bonded by the truncated LJ potential. (b) The LJ pair interaction potential between two bonded subunits of size $\sigma$ in the chain, for several values of attractive range, measured by $R_c$ and indicated by arrows in the plot.  (c) The contrast between a combined potential $W(r)$ felt by an inner subunit in the filament, bonded on both sides, and the end-subunit bonded by the regular LJ potential. (d) Scheme of the bond-bending force which opposes changes in the angle between two adjacent bonds by applying couples on each adjacent bond.}
\label{fig1}
\end{figure}

To model a non-covalently bonded filament we use a coarse-grained model of linear chains of Brownian particles (Fig.\ref{fig1}a) bonded by the truncated-shifted Lennard-Jones (LJ) potential,
\begin{equation}\label{1}
\frac{U_{LJ}}{k_BT} = \left\{ {\begin{array}{*{20}{c}}
{4\tilde \varepsilon [{{(\sigma /r)}^{12}} - {{(\sigma /r)}^6}] - {U_c},~~{\rm{ for }}~r < {R_c}}\\
{0, \qquad {\rm{ for }}~r \ge {R_c}}
\end{array}} \right.
\end{equation}
where $r$ is the distance between two neighbor proteins $i$ and $i+1$, $\sigma$  is the linear size of the monomer unit, and $U_c = 4\tilde \varepsilon [{(\sigma /{R_c})^{12}} - {(\sigma /{R_c})^6}]$. The parameter $\tilde \varepsilon  = \varepsilon /\{ 4[{(\sigma /{R_c})^{12}} - {(\sigma /{R_c})^6}] + 1\}$ is set to maintain a constant well depth equal to $\varepsilon$, independently of $R_{c}$.  The LJ potential is inherently anharmonic, except in the close proximity of its minimum. An alternative could be the Morse potential, and we have checked that the results do not change qualitatively with its use. Figure \ref{fig1}b explains what we mean by truncation: the attractive region stretches up to a distance $R_c$ (indicated by arrows in the plot and measured in terms of LJ length scale $\sigma$), while the depth of the potential well is kept independently fixed (measured by $\varepsilon$, in units of $k_BT$). The shorter the attraction range, the closer is the potential to its harmonic approximation. For all the data we use $\varepsilon=10$, which well approximates the strength of the most common physical interactions such as hydrogen bonds and hydrophobic attraction.

We also include in our analysis the local bending energy, in the form $\frac{1}{2} \sum_i K \, \theta_{i}^2$, where  $\theta_{i}$ is the angle between the directions of bonds from the particle $i$ to the preceding ($i-1$) and the subsequent ($i+1$) subunits. Figure~\ref{fig1}d illustrates the way this effect is implemented by imposing pairs of equal and opposite forces on the joining bonds, providing a net torque on the junction. It is the same algorithm that is used in, e.g. LAMMPS `angle-harmonic' system \cite{lammps}. The bending modulus $K$, {in units of $k_{B}T$,} is directly related to the persistence length of the filament via the standard expression $l_p \approx K \sigma/k_BT$.

The dynamics of the chain of subunits is governed by the overdamped Langevin equation,
\begin{equation}\label{A1}
\gamma \frac{{d{\bf{r}}}}{{dt}} =  - \nabla {V}({\bf{r}}) + {\bf{A}}(t)
\end{equation}
where $\mathbf{r}$ is the vector containing the positions of all molecules, $\gamma$ is the friction coefficient, the total potential force acting on a given particle, $- \nabla {V}$, has contribution from both the LJ and the bending couples, and
the Gaussian stochastic force defined such that $\langle {\bf{A}}(t)\rangle  = 0$ and
$\langle {A_i}(t){A_j}(t')\rangle  = 2{k_B}T \gamma \, {\delta_{ij}}\delta (t - t')$ , according to the fluctuation-dissipation theorem.
For numerical integration Eq. (\ref{A1}) is discretised in the form known as the Ermak-McCammon equation \cite{ermak1,ermak2}:
\begin{equation}\label{S2}
{\bf{r}}(t + \Delta t) = {\bf{r}}(t) - \frac{\nabla V({\bf{r}})}{\gamma}\Delta t + \Gamma \sqrt {\frac{{2 k_\mathrm{B}T}}{\gamma }\Delta t},
\end{equation}
where $\Gamma$ is randomly extracted from a normal distribution with zero average and unit standard deviation. The discrete time step is taken as $\Delta t = 5\cdot 10^{-5} \tau$, where the reduced time uint is defined as $\tau=\sigma^2/D$, and $D=k_\mathrm{B}T/\gamma$ is the diffusion coefficient. For a typical globular protein (e.g. Lysozyme), with diameter$\sigma \simeq 5$~nm and diffusion coefficient $D\simeq 10^{-10}$ m$^2$/s \cite{burne1}, we obtain $\tau\simeq 0.25$~$\mu$s. Therefore $\Delta t \simeq 13$~ps.
Each run is initialized with the equilibrium interparticle distance ${r_i} - {r_{i + 1}} = {2^{1/6}}\sigma $, as a straight chain (all $\theta_i=0$), corresponding to the minimum of all interaction potentials. A dissociation event is assumed to take place when one of the bonds exceeds the cut-off length  ($R_c$), i.e. $\left| {{\mathbf{r}_i} - {\mathbf{r}_{i + 1}}} \right| > {R_c}$, at which point the simulation is terminated and the location of the rupture recorded. The location of the rupture is recorded. To generate the probability distributions plotted in Figs. \ref{fig3} and \ref{fig4}, $N$ independent runs are performed and the normalised breakup probability is calculated as $P(s) = N(s)/N$ where $N(s)$ is the total number of recorded breakup events for the bond $s \in 1...n$. For most data we have reached $N \geq {10^4}$; since the runs are independent, the $N(s)$ are binomially (Bernoulli) distributed and the error bars are estimated as $\sqrt{{P(s)[1 - P(s)]/N}}$, which always stayed below 10\% of the value for $P(s)$.

\begin{figure} 
\includegraphics[width=.4\textwidth]{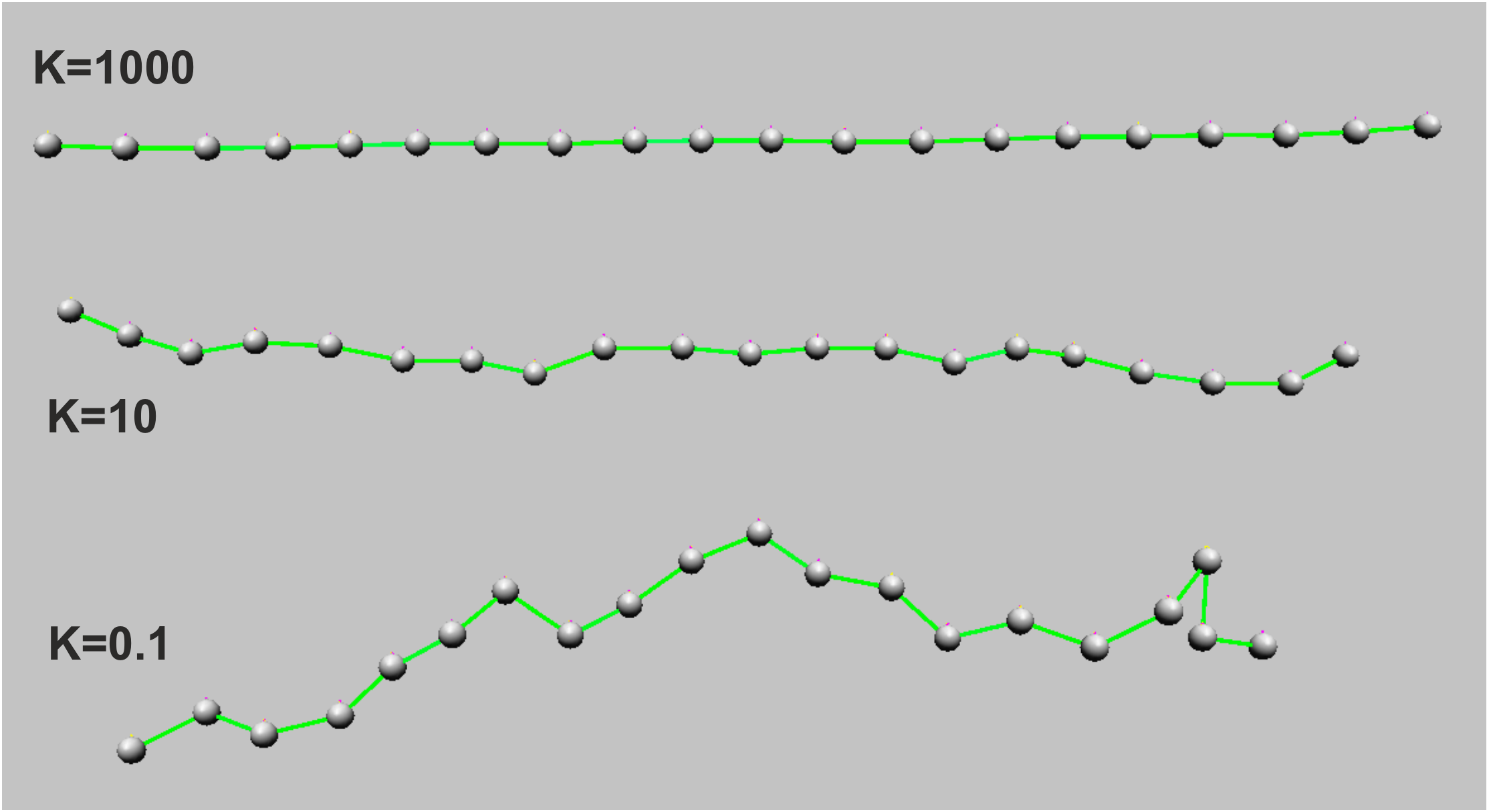}
\caption{An illustration of the role of bond-bending in the potential. The chains of $n=20$ subunits bonded with $\varepsilon = 10$ and $R_c=4$ were initialized from a straight conformation and allowed to fluctuate for $5000$ ts. The resulting snapshots, for each value of $K$ indicated on the image, show the effect of differing persistence length. }
\label{figK}
\end{figure}

\section{RESULTS}
\subsection*{Breakup statistics along the filament}
\begin{figure} 
\includegraphics[width=.42\textwidth]{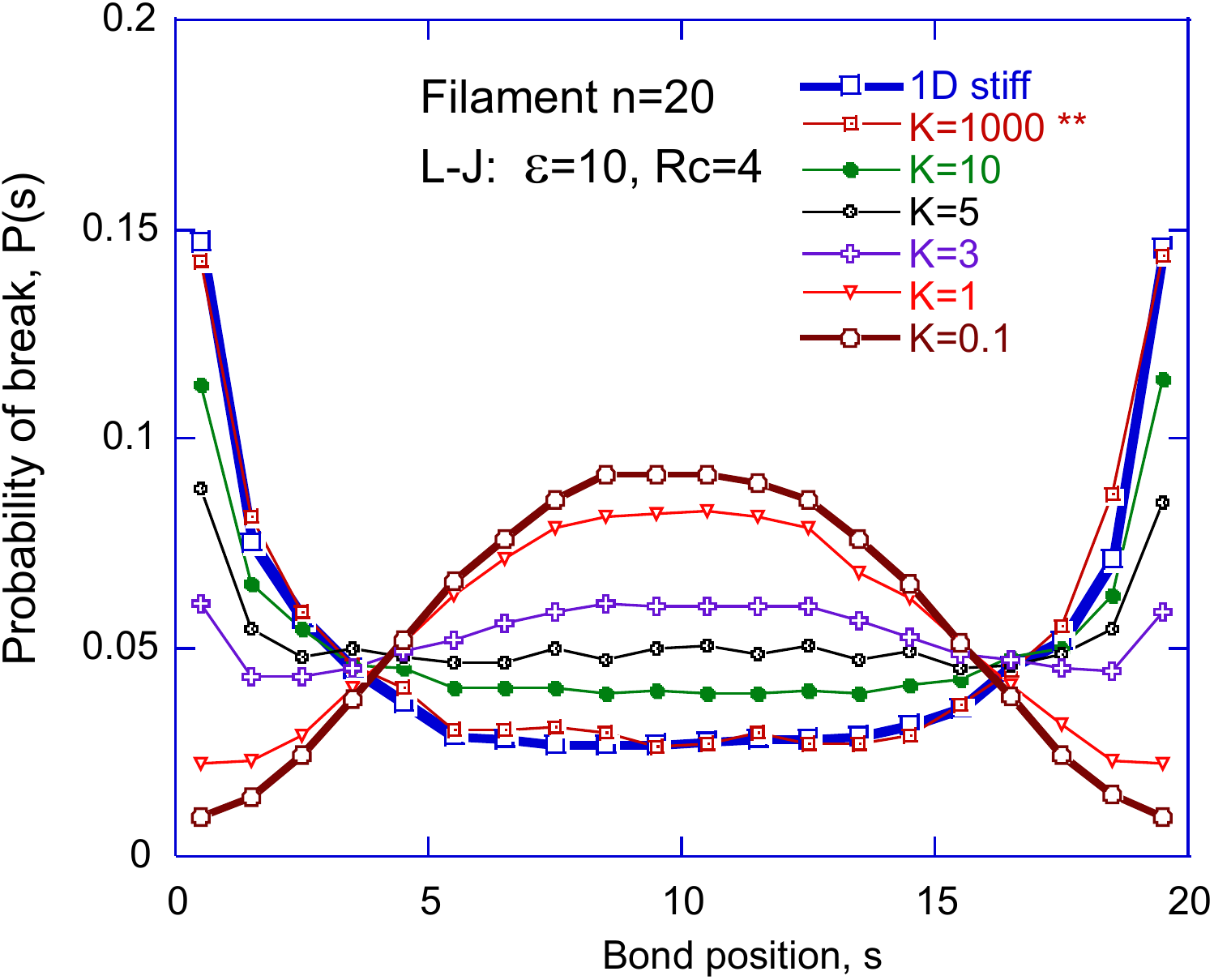}
\caption{The normalized probability of the first breakup as a function of the position along the filament for a chain $n=20$ and LJ parameters $\varepsilon=10$, $R_c=4$. The effect of changing bending stiffness (increasing persistence length) is evident: for the chain with essentially no bending penalty (lowest $K$) the distribution of fragment sizes is bell-shaped with a maximum in the middle of the middle of the filament.
Stiff chains, with a strong bond-bending penalty, instead, feature a nearly homogeneous, flat distribution of fragment sizes -- with an increasingly large increase of breakup rate at the ends. There is a broad range of semiflexible filaments that behave in exactly the same way: as ``stiff'' chains.}
\label{fig3}
\end{figure}

Figure \ref{fig3} shows the main result of our Brownian dynamics simulation: on increasing the bending stiffness of the filament, the highly inhomogeneous normalized probability $P(s)$ changes from a bell-shaped distribution peaked in the middle (reminding of the Hill model), to  a completely opposite shape, with a strong preference for single subunits to dissociate from the ends. 

The conclusion arising from this data is clear: there is a broad range of what one could collectively interpret as `stiff' filaments, for which the nature of bond-breaking statistics is exactly the same. These are with the bending stiffness of $K \gtrsim 1000$, and their behaviour does not differ from the last dataset in Fig.~\ref{fig3} (labelled `stiff'), corresponding to the strictly 1-dimensional filament where only the motions along the chain were permitted. For these stiff or semiflexible filaments there is a very strong preference to dissociate a single subunit from the chain ends, which does diminish for less symmetric potentials, as demonstrated by Fig. \ref{fig4} below. However, as the chain becomes increasingly flexible, the ratio of breaking rates at the ends and in the middle gradually reverses, and for a very flexible chain ($K = 0.1$ in the plot) the breakup probability resembles the prediction from the Hill model. One can qualitatively understand this effect: for a stiff filament (as shown in Figs. \ref{fig1}a and \ref{figK}), in order to develop a thermal fluctuation large enough to stretch a bond beyond $R_c$, a whole sequence of bonded particles has to move in a correlated fashion; this leads to an effectively harmonic potential acting on the middle particles, and diminishes their breaking rate very significantly.  On the other hand, as the particles in a flexible chain are free to move perpendicular to the bond axis, this coherent motion is not required and the bond breaking statistics is dominated by the single-bond equilibrium.

Most protein filaments are quite stiff. The F-actin has the quoted persistence length $l_p \sim 16 \, \mu$m \cite{actin1,actin2}, and the insulin amyloid filaments: $l_p \sim 4 \, \mu$m \cite{craig}. Interestingly, if one measures $l_p$ in the units of constituent protein size (as the parameter $\sigma$ in our case), these very different filaments all have $l_p$ between 3000 and 6000 units. We therefore choose the bending stiffness modulus $K=1000$ in all subsequent analysis, which is within the class of `stiff' filaments according to the data in Fig. \ref{fig3}.

\begin{figure} 
\includegraphics[width=.42\textwidth]{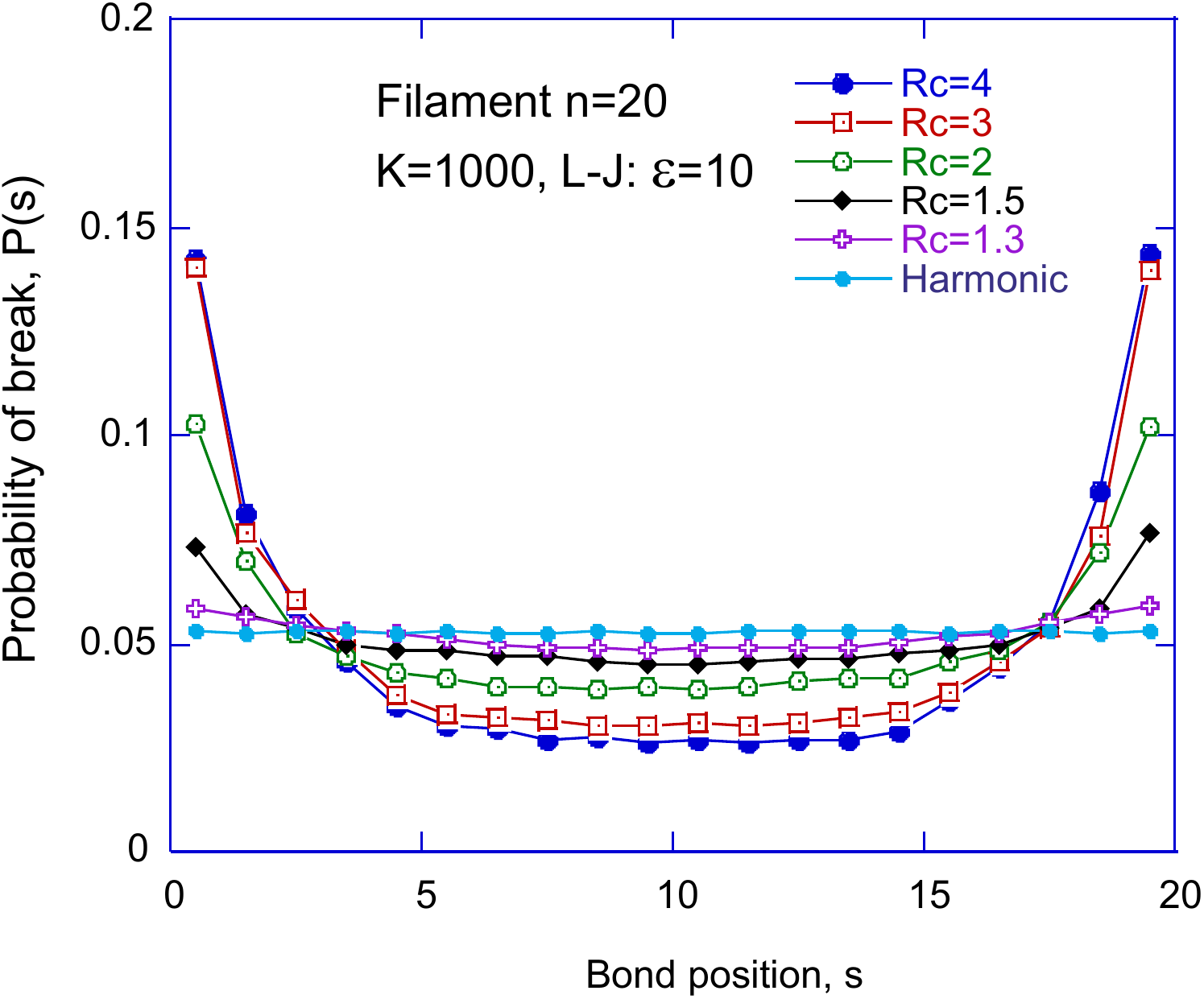}
\caption{The normalized probability of the first breakup event as a function of the position along the filament for a chain $n=20$ long, bonded by potentials with $\varepsilon =10$ and $K=1000$. The harmonic potential with the same depth has a uniform (homogeneous) probability of breakup, while at increasing $R_c$ (and the degree of potential asymmetry about the minimum) the ends of the chain are increasingly more prone to single-particle depolymerization. The ratio of the probability of breaking at the end (dissociation) to the fragmentation in the middle  $P_\mathrm{end}/P_\mathrm{mid}=5.46$ for $R_c=4$. }
\label{fig4}
\end{figure}

This distribution of breaking points along the chain is equivalent to the distribution of fragment sizes resulting upon breakup. Figure \ref{fig4} shows how this distribution depends on the nature of physical bond between subunits. As we have seen in the illustration, Fig. \ref{fig1}b, changing the cutoff distance $R_c$ while keeping the depth of the attractive potential well constant ($\varepsilon$) effectively alters the degree of potential asymmetry: the larger the $R_c$, the more asymmetric the potential is. We have also independently tested the breaking statistics in an explicitly harmonic potential of the same depth and curvature at the minimum.  In the limit of harmonic chain, we recover a completely uniform (flat) distribution of fragments, with a very high accuracy. This is in agreement with the theory of Lee~\cite{lee09}, who assumed harmonic bonds. On the other hand, Fig. \ref{fig4} clearly demonstrates that, with increasing asymmetry, the breakup probability $P(s)$ displays an increasingly strong preference for depolymerization from the ends. For the highly asymmetric (and also highly anharmonic) potential with $R_c=4$, the breakup probability of the outer bonds is over $5$ times larger than the one of the innermost bonds.

\begin{figure} 
\includegraphics[width=.42\textwidth]{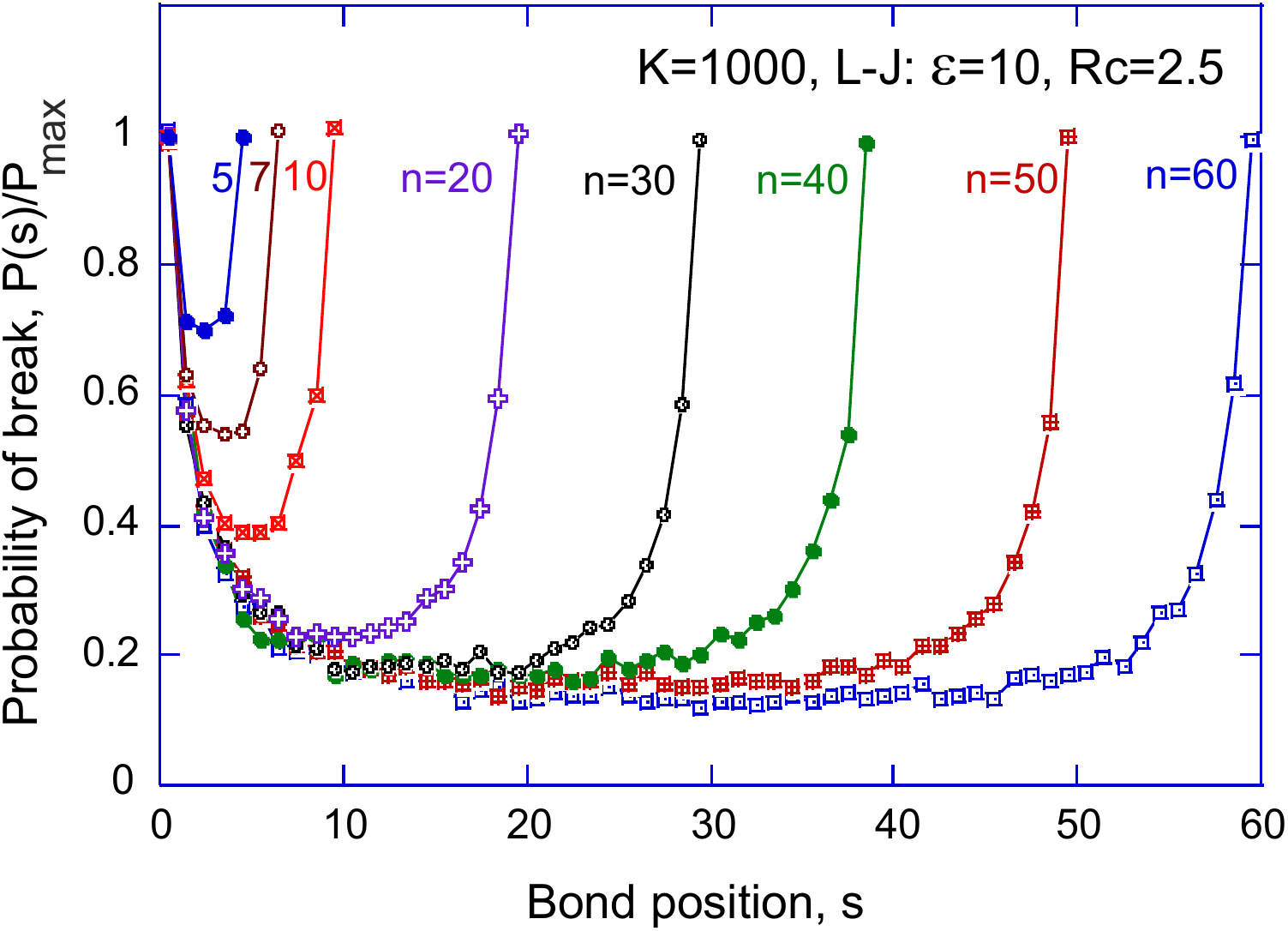}
\caption{Relative probability of the first breakup event upon varying the total length $N$, and the parameters of bonding potentials $\varepsilon = 10$, $R_c=2.5$ and $K=1000$.
The range of enhanced breakup probability $\Delta s $ at each end remains constant for all $n$.}
\label{fig5}
\end{figure}

Another important result is shown in Fig.\ref{fig5}, where for a given level of LJ potential depth and asymmetry, and stiff filament with $K=1000$, as usual, we study the effect of filament length (the total number of bonded subunits, $n$). It is more difficult to normalise the breakup probability $P(s)$ this time, because for longer filaments there are more and more `plateau values' of the constant (low) breakup frequency in the middle, which participate in the original normalisation by the total number of runs, $N$ (effectively uniformly suppressing the values of $P(s)$ and thus masking the characteristic ratio $P_\mathrm{end}/P_\mathrm{mid}$).  We therefore chose to scale all datasets by their maximum value of $P_\mathrm{end}$, such that the different curves are comparable. It is clear that with the filament increasing past $n=20$ there is no further change in the characteristic ratio $P_\mathrm{end}/P_\mathrm{mid}$ -- simply the region of `chain middle' becomes extended. Perhaps one may regard this as an effective confirmation of the Lee model \cite{lee09}, since for very long and very stiff filaments a very large middle section has an effectively harmonic bonding, and therefore uniform breakup rate.
It appears, the range of enhanced probability near the ends is relatively constant, $\Delta s \lesssim 10$. Shorter filaments have the middle region elevated simply because the two end-effects start overlapping.

 The finding that, for stiff filaments with asymmetric interaction potentials, the dissociation rate at the end can be substantially larger than the rate of fragmentation in the middle, may be important in the self-assembly kinetics of actin filaments \cite{oosawa75,pollard86}. There, and in many other cases, the tendency to depolymerize at the end is amplified by the presence of multiple bonds in the interior of the filament, due to the double-stranded helical structure in the case of actin.

\section{Probability of first breakup}
\begin{figure} 
\includegraphics[width=.4\textwidth]{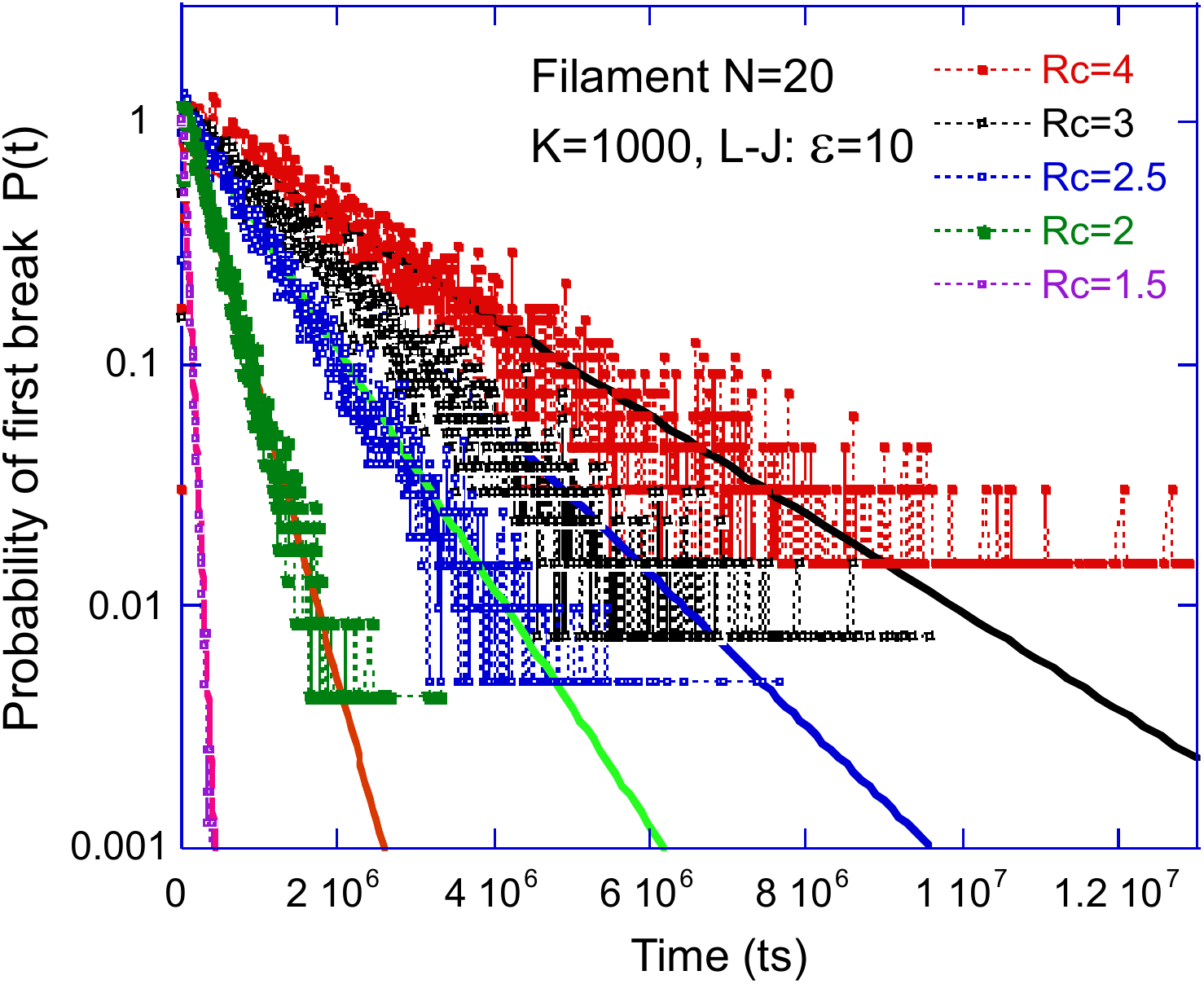}
\caption{The probability of first breakup of a filament of fixed $n=20$, normalized such that it is equal to unity at $t=0$, is plotted against simulation time measured in timesteps (ts). Different data sets represent the different attraction range $R_c$, which is our measure of potential asymmetry. The fitted lines are all simple exponentials, from which we extract the characteristic rate of the first breakup, $\lambda$. }
\label{fig6}
\end{figure}

In addition, we studied the probability of the first breakup (irrespective of its position along the filament), upon varying $R_c$ and the filament length $n$, for the case of a stiff filament (limit of large $K$) which also approximates the case of a 1D aggregate. From the results plotted in log-linear fashion in Fig.\ref{fig6} it is clear that the probability for the chain to fracture depends exponentially on time, $P(t) = \textrm{const} \cdot e^{-\lambda t}$, with a characteristic breakup time $\lambda^{-1}$  increasing upon increasing the attraction range $R_c$ (and the asymmetry of bonding potential with that). The average first-breakup time, irrespective of the location on the chain, is defined by $\lambda^{-1} = \int_0^\infty P(\tau) \tau \, \textrm{d}\tau$, upon normalizing $P(\tau) = \lambda \cdot e^{-\lambda t}$. This exponential dependence can be understood from the analysis of the many-particle Fokker-Planck equation,
\begin{equation}\label{2}
\frac{{\partial \rho ({\bf{r}},t)}}{{\partial t}} = {\hat L_s}\rho ({\bf{r}},t)
\end{equation}
with the Smoluchowski operator defined as~\cite{doi86,vankampen97}
\begin{equation}\label{3}
{\hat L_s}(...) = {D{\nabla _{{\textbf{r}}}} \cdot [{\nabla _{{\textbf{r}}}}(...) + \beta \nabla {U_{LJ}}(\textbf{r} )(...)]}
\end{equation}
acting on the many-particle probability density $\rho ({\bf{r}},t)$, where, in supervector notation, ${\bf{r}} = \{ {r_1},...,{r_n}\} $ is the set of interparticle coordinates. The probability as a function of time that all bonds remain within the cutoff at a time $t$, that is, the probability that the chain does not break within a time $t$, is given by
$Q (t) = \int_{ - \infty }^{{{\bf{R}}_c}} {\rho ({\bf{r}},t)d{\bf{r}}} $. We shall recall that, in supervector notation, the condition ${\bf{r}}<\bf{R}_c$ means that \textit{all} bond vectors (relative particle coordinates) in the chain are within an extension smaller than the cutoff $R_c$. Furthermore, ${U_{LJ}}(\textbf{r})$ represents the multi-dimensional potential energy landscape given by the superposition of the Lennard-Jones potentials acting on pairs of molecules.

The first passage/breakup time probability density is defined as the change of $Q(t)$ between the time $t$ and $t + dt$, and is thus given by $P(t) =  - dQ (t)/dt$. Combining these equations, with some manipulations (see e.g. Ref.~\cite{ebeling}), it is possible to show that the first-passage time probability density is exactly equal to
$P(t) =  - D{\left. {{\nabla _r}\rho ({\bf{r}},t)} \right|_{{{\bf{R}}_c}}}$. The mean first-breakup time is then defined as the first moment of the first-breakup time probability density,
${\lambda ^{ - 1}} = \int_0^\infty  {t \cdot P(t)dt}  = \int_0^\infty  {t \cdot [ - D{{\left. {{\nabla _r}\rho ({\bf{r}},t)} \right|}_{{{\bf{R}}_c}}}]dt} $, which is the same quantity as measured from the exponential fits in simulations. The exponential dependence on time can be understood from the analysis of the many-particle Fokker-Planck equation, Eqs. (\ref{2})-(\ref{3}). Its general solution is
$\rho ({\bf{r}},t) = \sum_{p} {{\phi _p}} ({\bf{r}}){e^{ - D{\lambda _p}t}}$, where $p$ labels the eigenfunctions $\phi _p$ and eigenvalues $\lambda_{p}$ of the many-body operator $\hat L_s$.

\begin{figure} 
\includegraphics[width=.32\textwidth]{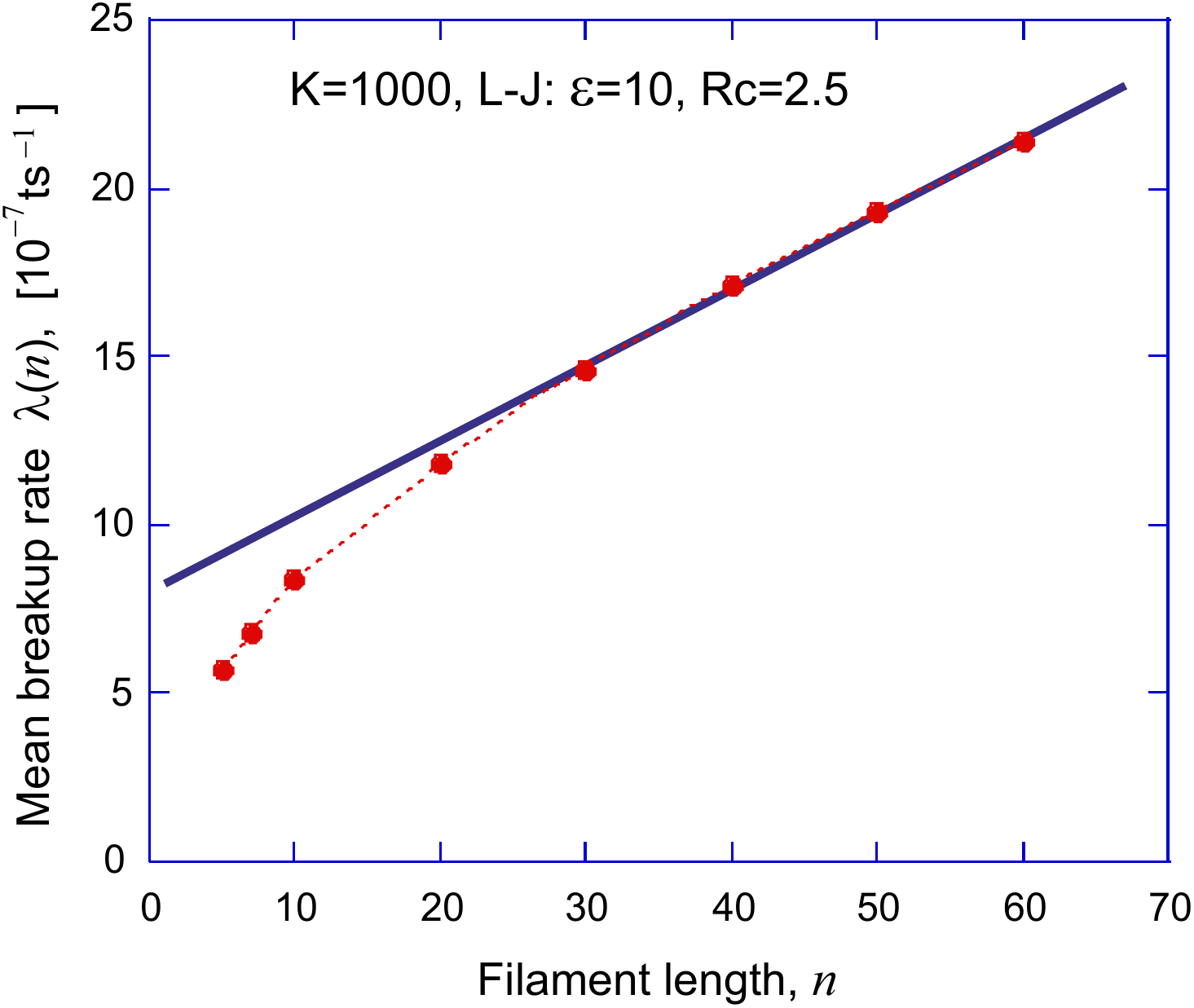}
\caption{The mean time of the filament breakup is plotted for different filament lengths, indicating an almost linear increase. In simple terms, taking from Fig.\ref{fig4} the probability to break in the middle as  $p_\textrm{mid} \approx 0.035$ and the one at the end as  $p_\textrm{end} \approx 0.145$, with $\Delta s \approx 6$ subunits from each end affected, the total rate can be estimated as $\lambda = (n-2\Delta s) p_\textrm{mid} +2\Delta s \, p_\textrm{end}$, which is the solid line in the plot with only a single fitting normalisation factor: $6.3 \cdot 10^{-7}\mathrm{ts}^{-1}$; the deviations at small $n$ are clearly due to the overlapping end effects (see Fig.\ref{fig5}).}
\label{fig7}
\end{figure}

According to the ground-state dominance principle, the time evolution for long filaments ($n \gg 1$) is dominated by the smallest non-zero eigenvalue $\lambda_{\mathrm{min}}$, such that, recalling the expression for $P(t)$, the time dependence of the first-breakup probability is given by
$P(t) =  - D{\left. {{\nabla _{\bf{r}}}\rho ({\bf{r}},t)} \right|_{{{\bf{R}}_c}}}\sim{e^{ - {\lambda _\mathrm{min}}t}}$. Hence the breakup probability is indeed exponential in time with a characteristic frequency-scale given by the smallest finite eigenvalue $\lambda_{\mathrm{min}}$  of the many-body operator $\hat L_s$. This result explains the exponential dependence on time of the breakup probability observed in the simulations in Fig. \ref{fig6}. Also, combining the expressions for $P(t)$ and for $\lambda^{-1}$, it is possible to show that $\lambda\approx\lambda_{\mathrm{min}}$, which confirms that the ground-state of the many-body Fokker-Planck equation indeed sets the time scale of breakup.

Furthermore, the rate $\lambda$  grows roughly linearly with the chain length $n$, which is demonstrated in Fig. \ref{fig7}. This particular dependence $\lambda \propto n$ arises because the number of escape attempts increases with the chain size. One can show by means of the standard supersymmetric transformation of the Fokker-Planck equation into the Schr\"{o}dinger equation~\cite{ebeling}, that $\lambda(n)$ is analogous to the quantum ground-state energy of an ensemble of $(n-1)$ bound states, and the ground state energy is extensive ($\propto n$) within the quasiparticle approximation~\cite{nozieres}.

\section{Discussion}

\subsection*{`Phase diagram' of first breakup locations}
\begin{figure} 
\includegraphics[width=.42\textwidth]{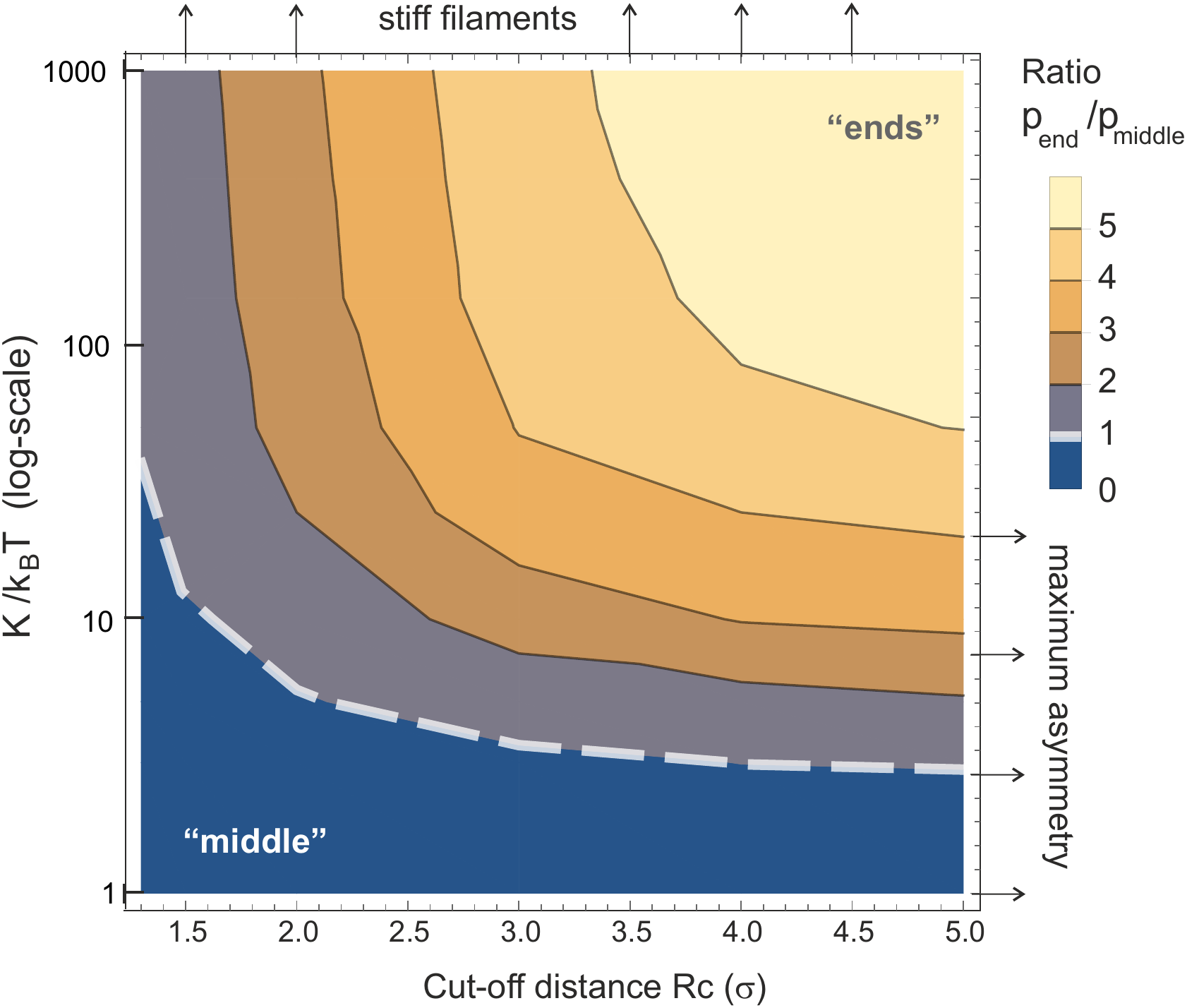}
\caption{Contour plot showing the ratio $P_{end}/P_{middle}$ as a function of filament stiffness $K$ and the asymmetry parameter $R_c$.  The bottom-left lagoon (dark) represents conditions where filaments break in the middle (bell-shaped distribution, according to Hill \cite{hill}), while the upper-right (light) region of this map  represents conditions where filaments break at the ends (the U-shaped distribution) and negligibly in the inner locations. Arrows on the top signify that these geodesic lines extrapolate towards $K\rightarrow \infty$. Arrows to the right indicate that there is little further change past $R_c = 4-5$. The dashed geodesic line marks the $P_{end}/P_{middle}=1$ condition, separating the regions of bell- and U-shape distributions. }
\label{fig8}
\end{figure}

We find a useful representation in a map that covers all of the $K$-$R_c$ parameter space to study how the location of first-breakup events along the filament changes upon varying both the stiffness $K$ and the cutoff or asymmetry $R_c$. 
The results can be represented as a contour plot for the ratio $P_{end}/P_{middle}$ as a function of $K$ and $R_c$. 
The contour plot is shown in Fig. \ref{fig8}. The bottom left corner, corresponding to flexible (low-$K$) filaments with short-ranged potential close to harmonic (low-$R_c$), represents conditions where the filament breaks in the middle and the fragment distribution is bell-shaped, in conformity with Hill's model predictions. Upon increasing both $K$ and $R_c$ at the same time, breakup in the middle becomes less favourable and the distribution tends to flatten out. Eventually, for very stiff filaments and asymmetric potentials with large $R_c$ the opposite limit of U-shaped fragment distributions with preferential breakup at the filament ends is recovered. This occurs in the top-left region of the map in Fig. \ref{fig8}. For symmetric binding potentials close to harmonic (low $R_c$: along the $K$ axis of the contour plot), the bell-shaped distribution persists longer upon increasing $K$, eventually transforming into a flat distribution $P_{end}/P_{middle}=1$ for stiff filaments. On the other side of the map, where $R_c$ is increased for flexible chains, the bell-shaped distribution persists for flexible chains up to $R_c \rightarrow\infty$ which corresponds to the LJ with no cutoff. 

In general, the most dramatic change in the breakup location and fragment-distribution shape occurs along the path of steepest ascent, defined as the path parallel to the gradient of the surface. Based on our results, the path of steepest ascent and most dramatic evolution in the breakup topology is approximately identified by the line $\log (K/k_BT) = (7/5) R_c/\sigma$. 

\subsection*{Bond-bending stiffness controls the filament breakup/recombination equilibrium}
In Figs. \ref{fig3} and \ref{fig4} we have shown that depending on the relative extent of bond-bending and central forces in the intermolecular interaction, the fragment size distribution can change from a U-shaped distribution in the limit of large bond-bending rigidity, to a bell-shaped distribution with opposite curvature in the limit of a purely central-force Lennard-Jones potential. Intermediate bending stiffness values yield distributions with shape in between the two limiting cases.

It is first important to understand the microscopic origin of this qualitative difference upon varying the bending stiffness in the intermolecular interaction. Since the flexible chain breakup statistics closely resembles the prediction of Hill~\cite{hill}, we take a similar approach and consider the fragment-size dependence of the breakup rate within a chemical equilibrium assumption and for the special simplifying case of harmonic bonds. We have checked that with harmonic bonds the same behaviour trend as in Fig. \ref{fig3} is reproduced, with the only difference that the $P(s)$ distribution for the stiff filament is flat (as indeed proven by Lee~\cite{lee09}) instead of U-shaped in the case of stiff filaments (as the last curve in Fig. \ref{fig4} shows). That is, the Hill-like bell-shaped $P(s)$ is the universal result for fully flexible chains.

The equilibrium constant for a dissociation reaction  $n \leftrightarrows n_{1} + n_{2}$ of a filament $n$ into two fragments $n_{1}, \, n_{2}$ takes the form: $K_\mathrm{eq}=V^{-1} Z({n_{1}}) Z({n_{2}})/Z({n})=K_{n1, n2}^{-}/K_{n1, n2}^{+}$, where $Z({n_{1}})$ is the partition function of fragment $n_1$. $K_{n1, n2}^{-}$ is the dissociation rate, while $K_{n1, n2}^{+}$ is the recombination rate of these two fragments. The latter can be estimated from the diffusion-controlled collision rate of two linear chains, upon accounting for the diffusion coefficient of the two chains (Kirkwood-Riseman approximation \cite{kirkwood}) and for the encounter efficiency of end-to-end collisions of the two chains. In this way, the size-dependence was found to be $K_{n1, n2}^{+} \propto (n_{2} \ln n_{1} + n_{1} \ln n_{2})/n_{1}n_{2}n$~\cite{hill}.
The size-dependence of the dissociation rate (and hence the fragment size-distribution) can be obtained by replacing this form for the association rate in the expression for the equilibrium constant, and upon evaluating the fragment-size dependence of the partition functions in the numerator of $K_\mathrm{eq}$.

From classical statistical mechanics, rigid-body translational degrees of freedom of the chain contribute to the partition function a factor $\sim n^{3/2}$, and rigid-body rotational degrees of freedom contribute an extra factor $\sim n^3$, since the overall mass of the filament is $\propto n$. Together these two factors give a partition function $\sim n^{9/2}$.
The vibrational contributions of the monomers in the chain factorise in the partition function, as for a chain of harmonic oscillators, resulting in standard factors of the type $\sim (k_BT/\hbar \omega)^n$, where $\omega$ is the Einstein frequency. Clearly these factors do not contribute to $K_\mathrm{eq}$ because the corresponding terms in the numerator and denominator cancel.

A full consideration of the normal modes of the linear chain with free ends, beyond the Einstein model, leads to an additional nontrivial size-dependence $\sim n^{-1/2}$, for vibrations of harmonic spheres in 1D, and to $\sim n^{-3/2}$ for vibrations in a flexible 3D chain~\cite{abraham,lothe}. In simple terms, upon increasing the chain length, more low-energy modes can be accommodated in the spectrum, which causes the partition function to decrease. The importance of this effect was first recognized by J. Frenkel~\cite{frenkel} in the context of nucleation phenomena. Hence with purely central-force interaction in 3D (flexible chain) the overall contribution is $\sim n^{9/2-3/2} = n^{3}$.
Akin to covalent bonds in molecular physics, the bending stiffness introduces additional degrees of freedom for rotations about the bond symmetry axis, which then leads to an overall dependence $\sim n^{9/2-3} = n^{3/2}$. One should note that with spheres and purely central-force bonds there is no such axis of symmetry for the rotations, and the three translational degrees of freedom per particle suffice to describe the vibrational behavior.
Including all these considerations, the dissociation rate will have a dependence on the fragment sizes given by
\begin{equation}\label{k1}
K_{n1, n2}^{-} \sim (n_{1}n_{2})^{x-1}(n_{2} \ln n_{1} + n_{1} \ln n_{2})/n.
\end{equation}
The exponent $x$, which collects all size-dependent contributions of the partition function, is different depending on whether the interaction is purely central-force, or has a bond-bending stiffness. For central forces, $x=3$, whereas with semiflexible or stiff chains one has $x=3/2$.
The leading contribution is then $\sim (n_{1}n_{2})^{2}$, with a pronounced bell-shape peaked in the middle for the exclusively central-force flexible chain, and $\sim (n_{1}n_{2})^{0.5}$, leading to a much flatter distribution for a chain with bond-bending penalty. {The fact that the slightly U-shaped distribution observed in simulations for stiff filaments is not recovered by this model should be attributed to the various approximations (Kirkwood-Riseman for chain diffusion, detailed balance, etc.) involved in the model, and also to the harmonic approximation of independent linear oscillators underlying the factorization of partition functions. }
This argument, however, explains, qualitatively, that a flatter distribution of fragments is to be expected in the presence of bond-bending, due to the additional rotational degrees of freedom about the stiff intermolecular bond symmetry axis, which is absent with purely central-force interactions.

\subsection*{Possible roles of electrostatics and temperature in amyloid fibril breakup}
We can briefly comment on the qualitative predictions of this model for the distribution of breakup fragments in realistic amyloid fibrils. {Realistic intermolecular forces which bind proteins in amyloid fibrils crucially depend on both electrostatics and temperature. We shall start considering the role of electrostatics first.}

{Electrostatic repulsion between two bound proteins in a filament is ubiquitous except for solutions at very high ionic strength. Electrostatic repulsion acts to ``lift up'' the bonding minimum, and it may also contribute an additional small energy barrier to the total interaction $U$, with a maximum $U_{max}$ co-existing or competing with the new lifted attractive minimum. We denote the new attractive minimum as $U_{min}^{*}<\epsilon$. Due to the fact that the electrostatic energy decreases with $r$, and the maximum is typically at $r>r_{min}$, the lifting up of the bonding minimum by the electrostatic repulsion is not entirely compensated by the energy barrier (the new maximum in $U$). Hence the total energy to be overcome for the particle to escape from the bonding minimum is $U_{max}-U_{min}^{*}<\epsilon$. 
This consideration points towards a role of electrostatics which promotes breakup, or at least, restructuring into a different morphology where the electrostatic energy density is reduced. This outcome of our analysis is compatible with recent experimental observations where an increased electrostatic repulsions (e.g. at lower ionic strengths) is responsible for fission or scission phenomena of larger compact aggregates into smaller and more anisotropic aggregates~\cite{Dehsorkhi,fodera1}.}

Our simulations show a crossover from a U-shaped fragment distribution into a bell-shaped distribution upon going from high values of bond-bending stiffness $K$ to lower values. In our simulations, $K$ is fixed and set independently of $T$, the latter being kept constant throughout at varying $K$. 
In reality, however, $K$ and $T$ may not be decoupled for a realistic model of amyloid fibrils. The reason is that the inter-protein bending stiffness $K$ originates, microscopically, from the strength of $\beta$-sheets which bind two adjacent proteins in the fibril. The mechanism is known: due to the planar, sheet-like, nature of two hydrogen-bonded $\beta$-sheets, there is an intrinsic bending resistance against sliding or rolling of the two proteins past each other. The same mechanism provides bending rigidity when two surfaces bonded by many anchored central-force springs are displaced tangentially apart. 
Upon increasing $T$, the hydrogen and hydrophobic bonds which keep the two $\beta$-sheets together start to dissociate, leading to lower bending stiffness and lower values of $K$. 

Hence, based on our simulation results, we can predict that the fragment distribution function of realistic amyloid fibrils should evolve from a U-shaped distribution at low temperature $T$, where the $\beta$-sheets of two adjacent proteins are tightly bound, into a bell-shaped distribution at higher $T$ where the $\beta$-sheet bonding becomes looser, which makes the bending stiffness $K$ decrease. This prediction seems to be confirmed by preliminary experiments~\cite{morbidelli}, and future work using ab-initio simulations should focus on identifying the relationship between $K$ and $T$, which controls the evolution of the fragment distribution with $T$. {In future research it will be important to combine all these effects into a general coarse-grained approach along the lines of~\cite{Knowles,Assenza}, to achieve a bottom-up description of realistic filaments and their size evolution. 
}

\subsection*{Anharmonicity controls depolymerization from the ends in stiff filaments}

When the bending rigidity of the chain is high, the probability of spontaneous bond breaking is flat when the bond potential is harmonic~\cite{lee09} -- yet it adopts a very distinct and very strongly biased U-shape when the /asymmetry of the potential increases (Fig.\ref{fig4}). How can we quantitatively explain why the asymmetry of interaction potential between any two bonded subunits leads to higher breakup rates at the chain ends, and much smaller breakup rates in the middle? For a high bending modulus one can treat the bond at the filament end as a classical diatomic molecule, and a subunit in the middle of the chain as the inner particle in a linear triatomic molecule. In the latter case, the combined potential $W$ felt by the particle in the middle is sketched in Fig.\ref{fig1}c. 

One would be tempted to explain the difference between the higher dissociation rate at the filament end and the lower one in the middle by referring to the overall lower energy (deeper potential well) felt by the particle in the middle sitting in the minimum of the combined potential $W(r)$. Applying a Kramers-type escape-rate argument would then lead to an Arrhenius dependence of the particle on the depth of the energy well and an overall large difference between the two rates.  However, such an approach cannot explain the observation that the rate is the same in the middle and at the end for the case of harmonic potential; in that case the same argument about $W$ applies hence one would expect a lower rate in the middle, which is not observed, in agreement with previous calculations~\cite{lee09}. What is different in the case of the harmonic potential, is the fact that the asymmetry of the bonding potential is removed for the particle at the end of the chain (while the subunits in the middle effectively experience the harmonic potential in both cases). 

It is in fact this asymmetry which facilitates dissociation at the termini of the chain, where less resistance is encountered by the particle escaping outwardly from the bound state. In order to verify that this is indeed the right physics, we also run a test simulation with a quartic potential $U\propto (r-r_{min})^{4}$, which is anharmonic yet fully symmetric about the minimum, just like the harmonic potential. Also in this case we found a completely flat distribution of fragments, as for the harmonic potential, which supports the proposed claim.

It is therefore the asymmetry, in the case of anharmonic potentials, which plays the major role in facilitating the preferential bond breakup at the chain ends. The explanation can be found in the different values of the mean thermal fluctuation from the equilibrium position (energy minimum) for the particle sitting in the asymmetric LJ potential at the chain end, and the particle moving in the more symmetric combined potential $W(r)$ in the middle of the chain. An analysis of the mean thermal fluctuation done long ago by J. Frenkel~\cite{frenkel}, shows that the mean thermal fluctuation of the particle feeling the anharmonic/asymmetric potential at the end is typically larger because of the shallower slope of the potential in the outward direction. For the particle in the middle, the situation is different because the combined potential $W(r)$ does not become shallower as the particle in the middle moves away from one of the two neighbours, due to the presence of the interaction with the other neighbour.

\section{CONCLUSIONS}
By means of Brownian dynamics simulations, we have shown that thermal breakup rates and breakup topology of model protein filaments (and other linear nanoparticle aggregates) are strongly affected by the presence of bond-bending stiffness in the interaction between subunits, and by the degree of asymmetry of the anharmonic binding potential. With stiff chains bonded by inter-particle forces with anharmonicity and asymmetry of the potential typical for intermolecular interaction potentials (van der Waals, hydrophobic attraction etc), we find a strongly preferential breakup at the chain ends, and an overall U-shaped fragment distribution.
In contrast, with purely central-force interactions between subunits, that is, fully flexible chains -- the fragment size distribution is bell-shaped, with a pronounced peak in the middle (symmetric breakup), and the lowest breakup rate is found at the ends of the chain.

While the preferential breakup at the end of stiff chains  (filament depolymerization) can be explained in terms of the larger thermal fluctuations at the chain-end associated with potential anharmonicity/asymmetry in a perfectly stiff quasi-1D chain model, the dramatic change of breakup topology upon varying the strength of bond-bending interaction is more subtle. In this case we found a tentative explanation upon considering the degrees of freedom associated with the vibrational partition function of the fragments. In general, breakup into two equal fragments is favoured with purely central-force bonds because the product of the partition functions of two fragments is maximised (which is intuitive if one considers that the classical partition function for rigid body motions increases strongly with the fragment size). The vibrational partition function, instead, decreases with fragment size because more low-energy modes can be accommodated in longer fragments. This effect becomes stronger in the case of bond-bending, where the total number of vibrational degrees of freedom is larger due to the rotation axis of the stiff bond. As a result of this compensation between the size dependencies of the vibrational and rigid-body partition functions, the size-dependence of fragmentation rate with bond-bending is much weaker compared to the central-force case.

Hence, we found some general laws which govern the fragmentation behavior of model linear aggregates, as a function of the relative importance of central-force and bond-bending interactions between subunits. These findings are important towards achieving a bottom-up control over the length and time-evolution of filament populations, both in biological problems (acting, amyloid fibrils etc.) and in nanoparticle self-assembly for photonic applications.

\begin{acknowledgments}
\noindent We are grateful for many discussions and input of
 T.P.J. Knowles, T. Michaels, C.M. Dobson and A. Bausch. This work has been supported by the
Ernest Oppenheimer Fellowship at Cambridge (AZ, LD) and by the Technische Universit\"{a}t M\"{u}nchen Institute for Advanced Study, funded by the German Excellence Initiative and the EU 7th Framework Programme under grant agreement nr. 291763 (AZ). LD also acknowledges the Marie Curie ITN-COMPLOIDS grant no. 234810.
\end{acknowledgments}


\begin{thebibliography}{99}

\bibitem{oosawa75} F. Oosawa, F. S. Asakura,  \emph{Thermodynamics of the Polymerization of Protein.} (Academic Press, 1975).
\bibitem{pollard86} T. D. Pollard,  Ann. Rev. Biochem. \textbf{55}, 987-1035 (1986).
\bibitem{dobson06} F. Chiti, C. M. Dobson,  Annu. Rev. Biochem. \textbf{75}, 333 (2006).
\bibitem{chandler05} D. Chandler,  Nature \textbf{437}, 640-647 (2005).
\bibitem{irback03} A. Irb\"ack, S. A.  J\'onsson, N. Linnemann, B. Linse, S. Wallin. Phys. Rev. Lett. \textbf{110}, 058101 (2013).
\bibitem{niranjan03} P. S. Niranjan, P. B. Yim, J. G. Forbes, S. C. Greer, J. Dudowicz, K. F. Freed, J. F. Douglas, J. Chem. Phys. \textbf{119}, 4070-4084 (2003).
\bibitem{adamcik10} J. Adamcik, J.-M. Jung, J. Flakowski, P. De Los Rios,	 G. Dietler, R. Mezzenga, Nature Nanotech. \textbf{5}, 423-428 (2010).
\bibitem{tanaka06} M. Tanaka, S. R. Collins, B. H. Toyama, J. S. Weissman, Nature \textbf{442}, 585 (2006).
\bibitem{wu1} A. Zaccone, D. Gentili, H. Wu, M. Morbidelli, J. Chem. Phys. 132, 134903 (2010).
\bibitem{wu2} H. Wu, A. Tsoutsoura, et al. Langmuir 26, 2761 (2010).
\bibitem{knowles09} T. P. J. Knowles, C. A. Waudby, G. L. Devlin, S. I. A. Cohen, A. Aguzzi, M. Vendruscolo, E. M. Terentjev, M. E. Welland, C. M. Dobson, Science \textbf{326}, 1533-1537 (2009).
\bibitem{fodera1} V. Fodera, A. Zaccone, M. Lattuada, A. M. Donald, Phys. Rev. Lett. \textbf{111}, 108105 (2013).
\bibitem{fodera2} L. Di Michele, E. Eiser, V. Fodera, J. Phys. Chem. Lett. \textbf{ 4}, 3158 (2012).
\bibitem{colloid09} S. Odenbach, ed. \emph{Colloidal Magnetic Fluids. Basics, Development and Applications of Ferrofluids}. (Berlin, Springer, 2009).
\bibitem{bonn98} B. Bonn, H. Kellay, M. Prochnow, K. Ben-Djiemiaa, J. Meunier,  Science \textbf{280}, 265-267 (1998).
\bibitem{DiMicheleJACS} L. DiMichele, et al. J. Am. Chem. Soc. 136, 6538 (2014). 
\bibitem{peyrard89} M. Peyrard, A. R. Bishop, Phys. Rev. Lett. \textbf{62}, 2755-2758 (1989).
\bibitem{mast13} C. B. Mast, S. Schink, U. Gerland, D. Braun, Proc. Natl. Acad. Sci. USA \textbf{110},  8030-8035 (2013).
\bibitem{lee09} C. F. Lee,  Phys. Rev. E \textbf{80}, 031134 (2009).
\bibitem{hill} T. L. Hill, Biophys. J. 44, 285 (1983).
\bibitem{fermi55} E. Fermi, J. R. Pasta, S. Ulam, Los Alamos Scientific Laboratory Report No. LA-1940, May 1955.
\bibitem{kruskal65} N. J. Zabusky, M. D. Kruskal, Phys. Rev. Lett. \textbf{15}, 240-243 (1965).
\bibitem{oliveira} F. A. Oliveira, P. L. Taylor, J. Chem.Phys. \textbf{101}, 10118 (1994).
\bibitem{vilgis} A. Ghosh, D. I. Dimitrov, V. G. Rostiashvili, A. Milchev, T. A. Vilgis, J. Chem. Phys. \textbf{132}, 204902 (2010).
\bibitem{zaccone12} A. Zaccone, E. M. Terentjev, Phys. Rev. Lett. \textbf{108}, 038302 (2012).
\bibitem{haenggi90} P. Haenggi, P. Talkner, M. Borkovec, Rev. Mod. Phys. \textbf{62}, 251-341 (1990).
\bibitem{wiggins13} F. A. L. Mauguiere, P. Collins, G. S. Ezra, S. Wiggins, J. Chem. Phys. \textbf{138}, 134118 (2013).
\bibitem{vilgis2} J. Paturej, A. Milchev, V. G. Rostiashvili, T. A. Vilgis, J. Chem. Phys. \textbf{134}, 224901 (2011).
\bibitem{lammps}S. J. Plimpton,  J. Comput. Phys. \textbf{117}, 1 (1995).
\bibitem{ermak1} D. L. Ermak, J. Chem. Phys. \textbf{62}, 4189 (1975).
\bibitem{ermak2} D. L. Ermak, J. A. McCammon, J. Chem. Phys. \textbf{69}, 1352 (1978).
\bibitem{burne1} D. Burne, S. Kim, Proc. Natl. Acad. Sci. U.S.A. \textbf{90}, 3835 (1993).
\bibitem{actin1}T. Yanagida, M. Nakase, K. Nishiyama, F. Oosawa, Nature, \textbf{307}, 58-60 (1984).
\bibitem{actin2}F. Gittes, B. Mickey, J. Nettleton, J. Howard,  J. Cell Biol. \textbf{120}, 923-934 (1993).
\bibitem{craig}T. P. J. Knowles, J. F. Smith, A. Craig, C. M. Dobson, M. E. Welland,
    Phys. Rev. Lett. \textbf{96}, 238301 (2006).
\bibitem{kirkwood}J. Riseman, J. G. Kirkwood, J. Chem. Phys. \textbf{18}, 512 (1950).
\bibitem{abraham} F. F. Abraham and J. Canosa, J. Chem. Phys. \textbf{50}, 1303 (1969).
\bibitem{lothe} J. Lothe and G. M. Pound, Phys. Rev. \textbf{182}, 339 (1969).
\bibitem{frenkel} J. Frenkel, \emph{Kinetic Theory of Liquids} (Dover, New York, 1946).
\bibitem{doi86} M. Doi, S. F. Edwards. \emph{The theory of polymer dynamics} (Oxford University Press, 1986).
\bibitem{vankampen97} N. G. van Kampen. \emph{Stochastic processes in physics and chemistry} (Elsevier, Amsterdam, 1997).
\bibitem{ebeling} W. Ebeling and I.M. Sokolov, \emph{Statistical Thermodynamics and Stochastic Theory of Nonequilibrium Systems} (World Scientific, Singapore, 2005).
\bibitem{nozieres} D. Pines and P. Nozieres, \emph{The Theory of Quantum Liquids}, vol. 1 (W.A. Benjamin, Reading Massachusetts, 1966).
\bibitem{Dehsorkhi} A. Dehsorkhi, V. Castelletto, I. W. Hamley, J. Adamcik, R. Mezzenga, Soft Matter 9, 6033-6036 (2013).
\bibitem{morbidelli} L. Nicoud, S. Lazzari, D. Balderas Barragan, and M. Morbidelli, preprint (2015). 
\bibitem{Knowles} T.P. Knowles, et al. Phys. Rev. Lett. 109, 158101 (2012).
\bibitem{Assenza} S. Assenza, J. Adamcik, R. Mezzenga, P. De Los Rios, Phys. Rev. Lett. 113, 268103 (2014). 

\end{thebibliography}
\end{document}